\documentclass[conference]{IEEEtran}
\pdfoutput=1
\usepackage{url}
\usepackage{graphicx}
\usepackage{amsmath}
\usepackage{amsmath,amssymb,amsfonts}
\usepackage{algorithmic}
\usepackage[ruled,vlined]{algorithm2e}
\usepackage{multirow}

\ifCLASSINFOpdf
\else
\fi
\IEEEoverridecommandlockouts
\begin{document}
%
\title{On Hardware-Aware Design and Optimization of Edge Intelligence\thanks{Huai, Kong, and Luo contribute equally and are ordered alphabetically.}\thanks{\copyright~2023 IEEE. Personal use of this material is permitted. Permission from IEEE must be obtained for all other uses, in any current or future media, including reprinting/republishing this material for advertising or promotional purposes, creating new collective works, for resale or redistribution to servers or lists, or reuse of any copyrighted component of this work in other works. This is the author's accepted version of the article published in IEEE Design \& Test, vol. 40, no. 6, pp. 149-162, December 2023, DOI: 10.1109/MDAT.2023.3307558.}}

\author{
\IEEEauthorblockN{
Shuo Huai\IEEEauthorrefmark{1}\IEEEauthorrefmark{2}, 
Hao Kong\IEEEauthorrefmark{1}\IEEEauthorrefmark{2}, 
Xiangzhong Luo\IEEEauthorrefmark{1}, 
Di Liu\IEEEauthorrefmark{3}, 
Ravi Subramaniam\IEEEauthorrefmark{4},\\ 
Christian Makaya\IEEEauthorrefmark{4},
Qian  Lin\IEEEauthorrefmark{4}, 
Weichen Liu\IEEEauthorrefmark{1}}
\IEEEauthorblockA{\IEEEauthorrefmark{1}School of Computer Science and Engineering, Nanyang Technological University, Singapore}
\IEEEauthorblockA{\IEEEauthorrefmark{2}HP-NTU Digital Manufacturing Corporate Lab, Nanyang Technological University, Singapore}
\IEEEauthorblockA{\IEEEauthorrefmark{3}Department of Computer Science, Norwegian University of Science and Technology, Norway}
\IEEEauthorblockA{\IEEEauthorrefmark{4}HP Inc., Palo Alto, California, USA}
}


%


\maketitle

\begin{abstract}
Edge intelligence systems, the intersection of edge computing and artificial intelligence (AI), are pushing the frontier of AI applications. However, the complexity of deep learning models and heterogeneity of edge devices make the design of edge intelligence
systems a challenging task. Hardware-agnostic methods face some limitations when implementing edge systems. Thus, hardware-aware methods are attracting more attention recently. In this paper, we present our recent endeavors in hardware-aware design and optimization for edge intelligence. We delve into techniques such as model compression and neural architecture search to achieve efficient and effective system designs. We also discuss some challenges in hardware-aware paradigm.
\end{abstract}


%
\IEEEpeerreviewmaketitle

\section{Introduction}

The rule of thumb when designing deep learning models is \textit{the higher complexity, the better accuracy}. 
This rule almost applies to all deep learning models, such as convolutional neural networks (CNNs) and large language models (LLMs) which are widely exposed to public recently due to the breakthrough success of ChatGPT. 
Meanwhile, models are expected to be implemented at the \textit{edge} close to data so that some computation can be processed locally and expensive communications can be avoided \cite{deng2020edge,liu2022bringing}. 
Recently, new efforts even strive to execute LLMs offline on mobile devices without accessing powerful servers\footnote{\url{https://mlc.ai/mlc-llm/}}. 

To expedite complex models on the edge, numerous edge accelerators were devised in past years and more are expected in the near future \cite{hennessy2019new}. 
Due to high data parallelism and relatively simple operations, mainly addition and multiplication, emerging accelerators feature many simple processing units and deploy an advanced communication infrastructure, e.g., network on chips \cite{nabavinejad2020overview}, to connect all processing units \cite{deng2020model}. 
Model training and inference on accelerators involve loading a huge amount of model's weights to processing units and transferring intermediate activation layer by layer via communication infrastructure. 
Accelerators deploy diverse design paradigms with different processing units and communication infrastructure. 
As a result, this leads to significant performance variations of the same model on different platforms \cite{wu2019machine} and in turns propels to exploit hardware-aware design and optimization \cite{li2021hw}. 

To unleash the full potential of an edge intelligence system, a model should fully utilize the underlying hardware resources so that the accuracy may be maximized while the required performance is guaranteed. 
Hence, the model should be customized for the target hardware platform. 
However, there is a huge design space to explore when designing an edge intelligence system, diverse hardware accelerators, various intelligent applications with different datasets, and different performance requirements. 
There are two ways to achieve this goal: \textit{design} or \textit{optimization}.
\begin{itemize}
    \item When an extant model is too complicated to implement on an edge platform, compressing the model which is designed in a hardware-agnostic fashion can tailor the compressed model for the edge.
    \item We also can design a new and lightweight model to fit an edge platform, where the optimal accuracy can be retained and the required performance can be guaranteed.
\end{itemize}
Both methods should be conducted in a hardware-aware fashion to guarantee that the optimized or newly designed model can achieve or satisfy the expected performance requirement. 
In this paper, we discuss our recent endeavors to design and optimize edge intelligence systems in a hardware-aware manner and we also outline some unaddressed challenges. In our works, we take CNNs as the main target. 

The remainder of this paper is organized as follows: 
Section \ref{section:back} discusses the modelling techniques needed for hardware-aware methods. 
Section \ref{section:methods} briefly presents three efforts we recently made in designing and optimizing edge intelligence systems. 
Section \ref{section:challenges} further discusses some challenges that can be addressed in this area. 
Section \ref{section:concl} concludes this paper.
\section{Hardware and CNN Modelling}
\label{section:back}
Designing and optimizing edge intelligence is a non-trivial task due to a huge design space, spanning from model design or selection to training configurations, optimization methods, and hyper-parameters tuning. 
It is prohibitively expensive to train each design point from scratch and evaluate it on the target edge platform. 
This not only incurs a significant training expenditure but also environmental sustainability issue caused by high power consumption of servers. 
Thus, a more common way to evaluate the quality of a specific design is to model the target hardware, so that the performance or energy consumption of one design  can be quickly and approximately obtained using a hardware model. 
In some cases, a model that can predict accuracy of an input CNN model with different configurations is also desired, e.g., model compression and scaling, where such accuracy model can guide the optimization. 
Hardware and accuracy models significantly narrow the design space and boost the search procedure. 
In this section, we discuss the hardware modelling technique as well as the accuracy modelling technique used in our methods. 
\subsection{Hardware Modelling}
\label{model:hardware} 
The needs of hardware modelling are twofold: \textit{The inaccurate proxy metrics}--The direct performance metrics of a model are latency, throughput, or energy consumption. These metrics can be obtained by evaluating models on the target platform. 
It is difficult to always evaluate models on the target hardware, given a huge design space. 
Hence, many works tend to use some \textit{proxy} metrics, like FLOPs and MACs, to represent real performance of a CNN model. 
The application of proxy metrics is based on an assumption that there is a straight mapping from real metrics to proxy metrics. 
Unfortunately, this assumption does not always hold, and reduction on proxy metrics cannot translate into real reduction in terms of latency or energy \cite{luo2022lightnas,nabavinejad2020overview}. 
The second reason to have a hardware model is that some literature reveals that hardware architectures have considerable impact on a model's performance \cite{deng2020model}. 
Directly measuring the performance on the target device is favorable but infeasible due to the aforementioned design space and diverse ecosystems used by different edge hardware. 
Such diversity usually requires a conversion from a general framework, like PyTorch, to a specific format only supported by a specific hardware. 
As a result, A large design space accumulates to a non-negligible conversion overhead \cite{kong2021edlab}. 
The two reasons together propel the development of hardware modelling. 

\begin{figure}
    \centering
    \includegraphics[width=\columnwidth]{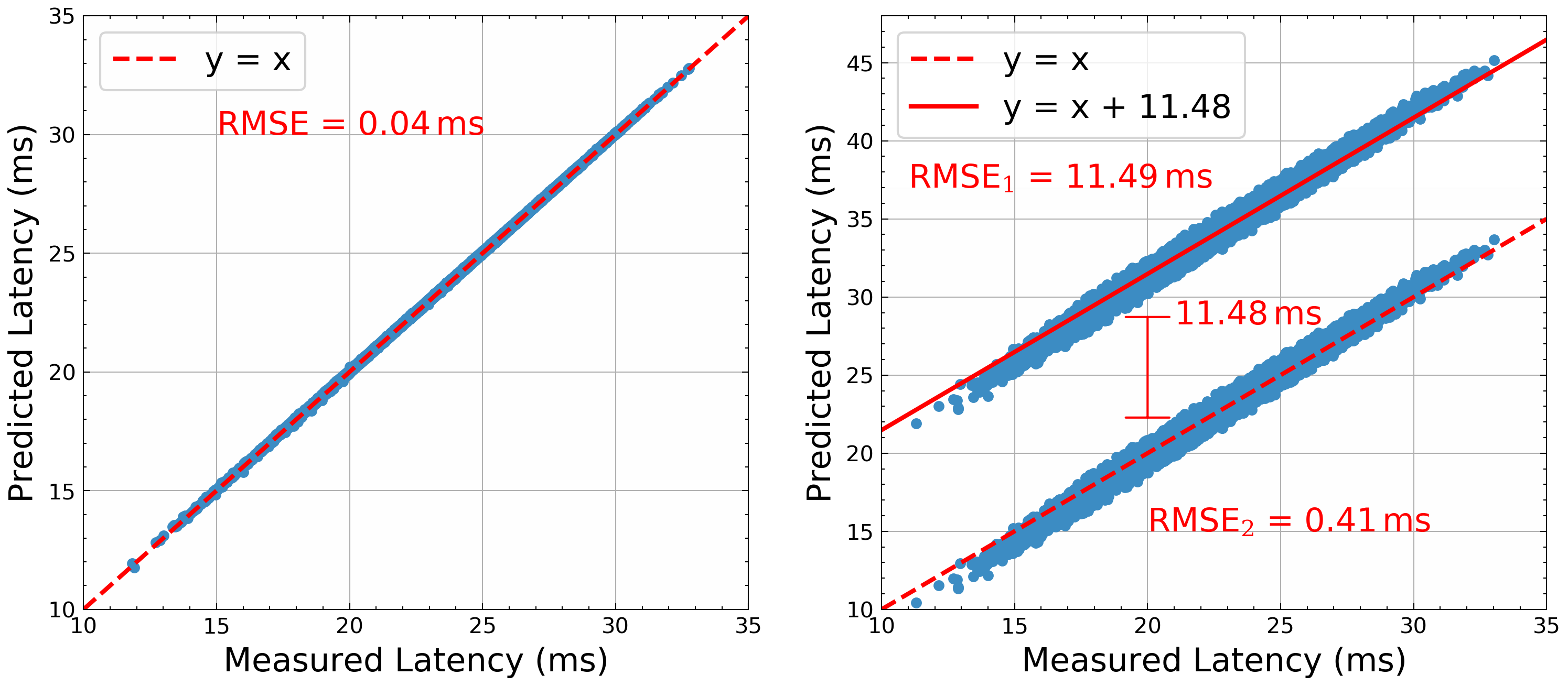}
    \caption{The comparison between MLP and LUT \cite{luo2022lightnas}.}
    \label{fig:MLP-LUT}
\end{figure}

Since CNNs have relatively simple operations, mainly addtion and multiplication, some strive to use look-up tables (LUTs) to model performance of CNNs. However, LUTs are unable to capture the complex data transmission occurred within the target hardware, thereby leading to imprecision when a model changes \cite{luo2022lightnas}. 
To address the inferiority of LUTs, others, including our works, start to use machine learning (ML) based methods to model performance of CNNs on a hardware. 
ML-based methods require to collect some operational data and train an ML model with the data. 
ML models used to predict results can be simple multi-layer perceptron (MLP) or more complex graph neural networks. We found from our experiments, a simple MLP is already able to achieve a relatively good result in terms of prediction accuracy. Hence, in our methods, we mainly use MLP as means to model hardware performance. Figure \ref{fig:MLP-LUT} shows one result we obtained for Nvidia Jetson AGX Xavier, where the left figure shows the results of our MLP-based modelling and the right figure shows the comparison between our MLP modelling ($\text{RSME}_2=0.41ms$) and a LUT-based modelling ($\text{RSME}_1=11.49ms$). 

\subsection{CNN Modelling}
\label{model:cnn}
CNN modelling aims at modelling a model's accuracy. It is required, because the accuracy of a model can only be evaluated after the model has been fully trained. Training is known to be computation-intensive, especially for some complex dataset. 
Some model-level modifications, like pruning  and compression, need to re-train the model to retain its accuracy, and this tedious procedure and high cost prohibits the possibility of exploring a large design space. 
As a result, similar to hardware modelling, some works propose to model the accuracy of different configurations of a CNN architecture in order to shrink the large design space and speed up the exploration procedure. 
Thanks to some interesting observations from the network design space exploration \cite{radosavovic2020designing}, CNN modelling only needs to sample a small amount of models with different configurations and train them with a small number of epochs. 
We only use a CNN modelling in multi-dimensional compression discussed in Section \ref{smartsissor}, where it facilitates the quick exploration of a large design space. 
However, we envision that other design and optimization methods may also benefit from a good CNN modelling.
\section{Hardware-Aware Design and Optimization}
\label{section:methods}
In this section, we present three of our recent works in hardware-aware design and optimization. We start with the widely-studied model compression.

\subsection{ZeroBN}
\label{subsection:zerobn-methodology}

The first work presented in this section is a pruning method, namely \textit{ZeroBN}. 
\textit{ZeroBN} is motivated by two flaws of existing pruning methods. 
Similar to what we discuss in Section \ref{model:hardware}, existing methods deploy proxy metrics to guide their pruning procedures. 
This is difficult for a pruning method to achieve a target latency which is paramount for many latency-sensitive applications.
To this end, these pruning methods have to repeat a three-stage procedure, \textit{pre-training}, \textit{pruning}, and \textit{fine-tuning}, to guarantee the required latency, while maximizing the accuracy of the pruned model. 
This drives us to think about how we can prune a model to satisfy its latency requirement in an efficient way.

\begin{figure*}[!ht]
    \centering
        \setlength{\abovecaptionskip}{0pt}
    \includegraphics[width=0.98\textwidth]{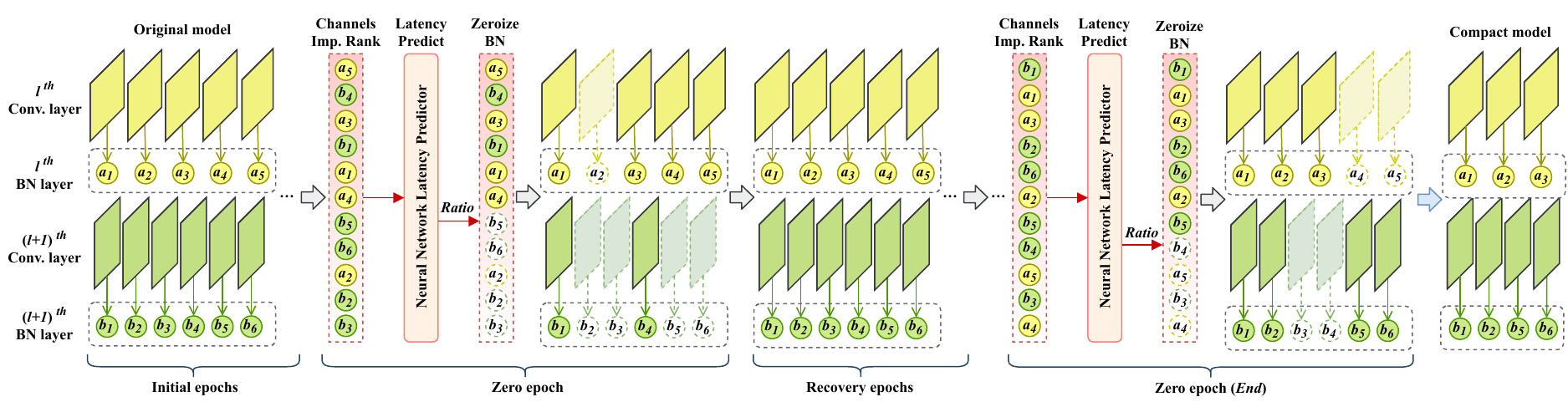}
    \caption{The process of ZeroBN \cite{huai2021zerobn}.}
    \label{fig:zerobn}
\end{figure*}

\textit{ZeroBN} is our answer for this question \cite{huai2021zerobn}.
\textit{ZeroBN} is a learning-based pruning method to directly learn a compact model that can satisfy the latency requirement. 
The main advantage of ZeroBN is that it is a \textit{one-shot learning process}, i.e., by having a normal training-like process, it can derive a pruned model with a competitive accuracy and latency guarantee. 
The overview of \textit{ZeroBN} is shown in Fig. \ref{fig:zerobn}, where it takes as input a large redundant CNN without training and outputs a compact and well-trained model that satisfies its latency constraint. 

\textit{ZeroBN} divides a traditional training period into three phases to implement the efficient pruning method, where the three phases are 1) \textbf{Initial Training} (IT); 2) \textbf{Zero Training} (ZT); and 3) \textbf{Recovery Training} (RT). The three phases are similar to the three stages in normal pruning methods, pre-training, pruning, and retraining. However, \textit{ZeroBN} combines all them into one training period. 

\textbf{IT} phase is equivalent to the pre-training in normal pruning methods, but \textbf{IT} is only conducted for several epochs at the beginning of the whole procedure. \textbf{IT} aims to obtain some initially trained weights for the input model so that we can evaluate the importance of different channels in later phases. 
One significant difference between \textit{ZeroBN} and others in terms of training is that we adopt sparsity training in the whole \textit{ZeroBN} process to impose sparsity regularization on scaling factors which are used to identify redundant channels and prune the model. 

After several epochs of \textbf{IT}, \textit{ZeroBN} proceeds to two iterative phases: \textbf{ZT} and \textbf{RT}. 
During these two phases, \textbf{ZT} will \textit{"soft-prune"} the input model by temporarily excluding some redundant channels based on their scaling factors and a compression ratio which is determined according to the latency constraint and a latency prediction model. 
The purpose of \textbf{ZT} is to derive a compressed model, train and evaluate it. 
During \textbf{ZT}, a latency predictor is integrated to efficiently calculate a compression ratio which can guarantee the compressed model meet its latency requirement. 
The latency predictor is the same to what we introduced in Section \ref{model:hardware}.
\textit{ZeroBN} exploits a global pruning method, i.e., we rank all channels within the model and determine redundant channels based on the global ranking and the computed compression ratio.

\textbf{ZT} trains the "compressed" model for a certain number of epochs, and then \textbf{RT} restores the "compressed" model to the full model and trains it for a number of epochs. 
The rationale behind \textbf{RT} is that
although the compressed model generated by a \textbf{ZT} phase can satisfy its latency constraint, it may be not the optimal model in terms of accuracy due to insufficient training and different training batches. 
Thus, after a \textbf{ZT} phase, we introduce a \textbf{RT} phase that restores the compressed model back to the full scale.
\textbf{RT} can avoid a \textbf{ZT} phase from ending up with a suboptimal model, thereby giving \textit{ZeroBN} a chance to learn a better compressed model. 
\textbf{ZT} and \textbf{RT} are interleaved after the \textbf{IT} phase, and the whole procedure ends with a \textbf{ZT} phase that generates the final compact model with latency guarantee. 

\begin{figure}[!t]
    \centering
    \setlength{\abovecaptionskip}{0pt}
    \includegraphics[width=0.48\textwidth]{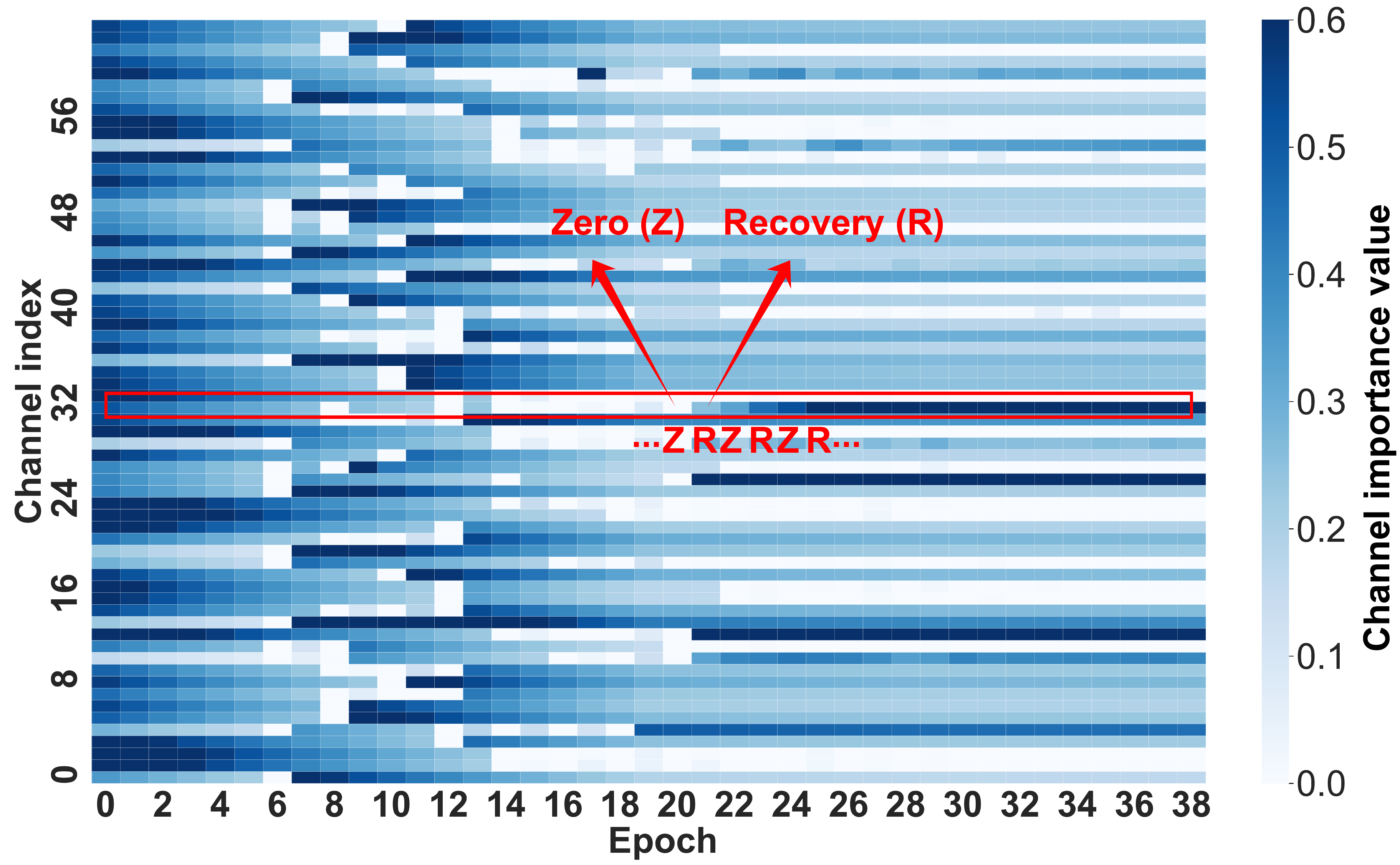}
    \caption{Channel importance changes during ZeroBN \cite{huai2021zerobn}.}
    \label{fig:zerobn_channel}
\end{figure}

Figure \ref{fig:zerobn_channel} shows the changes in the importance of all channels during the \textit{ZeroBN} process on an example model, where the number of epochs for \textbf{IT} is $6$, the compression ratio $\delta$ is set to a constant $0.3$, and \textbf{ZT} and \textbf{RT} both take 2 epochs for training. From this figure, we can see how channels' importance is changed over the training procedure, where some channels exhibit consistent importance and others vary significantly. 
This justifies the necessity of \textbf{RT}.

\textit{ZeroBN} has the same number of epochs as a normal training, where we just split the training epochs into the three parts introduced above. Thus, it does not add any extra overhead in terms of training and pruning. From the experiments \cite{huai2021zerobn}, we found that \textit{ZeroBN} is an efficient method to design compact models for diverse hardware. It has been open-sourced at \url{https://github.com/HPInc/ZeroBN}. We also have explored the potential of such learning-based pruning method in collaborated learning paradigm \cite{huai2022collate}. 


\subsection{SmartScissor}
\label{smartsissor}
\textit{ZeroBN} provides a means to compress a redundant model for edge systems, but it only considers one dimension of CNN models, i.e., the width (the number of channels per layer). 
Besides the width, there are two other dimensions which can affect the complexity of a CNN model, i.e., the depth (the number of layers) and the resolution (the size of inputs). 
When it comes to edge systems, we have to consider two factors, \textit{model complexity} and \textit{computational cost}. Model complexity reflects the number of parameters and weights, or MACs/FLOPs in a model, while computational cost shows the execution complexity, mainly intermediate activations generated by a model during its inference. 
The implementation of complex models on memory-limited edge systems is hindered by high model complexity and computational cost. 
Compressing width or depth is able to reduce both model complexity and computational cost, while compressing input size can significantly reduce computational cost. 
For instance, as the MACs of a CNN reduce quadratically with respect to the image input size, many works resize the input images to a smaller resolution (e.g., 112$\times$112) to reduce the computational cost \cite{kong2022smart}.

The three dimensions of CNN models present an opportunity to compress a complex model in a joint way, instead of only compressing width as \textit{ZeroBN}. EfficientNet \cite{tan2019efficientnet} is the pioneer work to explore the joint compression and finds such joint compression can achieve better accuracy with lower model complexity. 
However, there is no free lunch, and this opportunity also poses some challenges. Since three dimensions can be tuned, it exhibits a huge design space. 
As discussed before, once a small modification is applied to a model, the modified model has to be re-trained to retain accuracy, thereby resulting in high overhead. 
Thus, a more feasible way as did in EfficientNet \cite{tan2019efficientnet} is to have a compound shrinking\footnote{We use shrinking to refer to compression throughout this session for \textit{SmartScissor}, because we have used shrinking in the original framework.}, i.e., finding one coefficient for all three dimensions and using it to scale the three dimensions simultaneously. Nevertheless, finding the optimal compound coefficient is a challenging task that still needs significant effort. 

\begin{figure}
    \centering
    \includegraphics[width=0.45\textwidth]{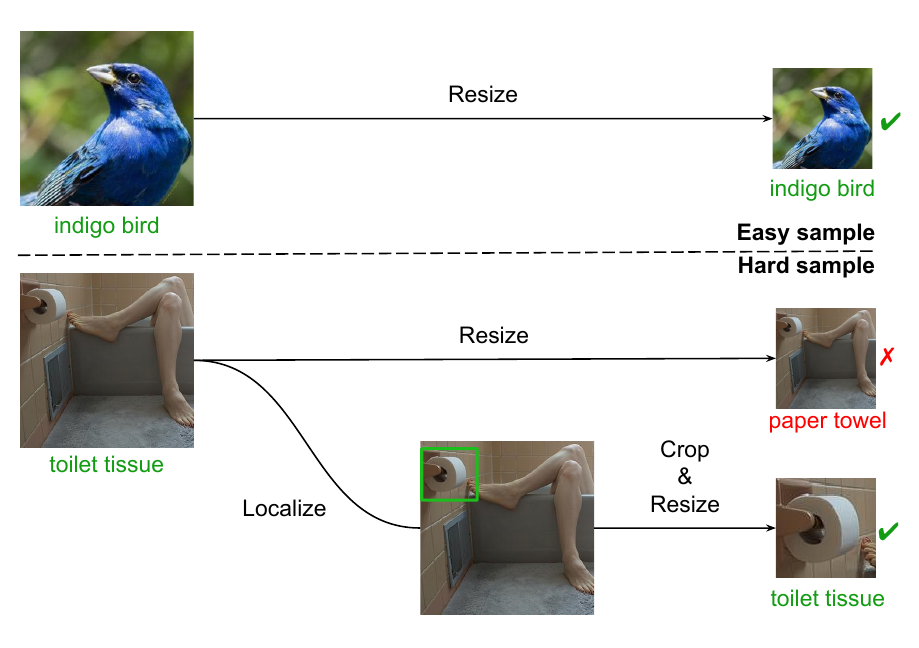}
    \caption{The prediction results of our pretrained ResNet-50 model. For easy samples, the network can still generate correct predictions at a small resolution (e.g. 112 $\times$ 112 for ImageNet). For hard samples, simply resizing the images to a small resolution can lead to misclassification, while the dynamic cropping strategy can correctly classify hard samples at the small resolution \cite{kong2022smart}.}
    \label{fig:motivation}
\end{figure}

When searching for the optimal compound coefficient, some additional attention must be paid to resolution. 
Different images demonstrate varying levels of classification difficulties. 
As demonstrated in Fig. \ref{fig:motivation}, easy samples with clear foreground can be correctly recognized even at a small resolution. 
For hard samples, as the foreground object only occupies a small portion of the whole image, directly shrinking the image to a small resolution will lose the details of the object, leading to a misprediction. 
Nevertheless, if we can crop the foreground object for inference, even hard samples can be correctly classified at a lower resolution. 
However, existing image preprocessing methods, e.g., ResizedCenterCrop (RCC), crop all images in a static manner and cannot achieve such instance-aware fine cropping. 
The efficiency of compounding shrinking and the observation we obtained from image resizing shown in Fig. \ref{fig:motivation} motivate our work, namely \textit{SmartScissor}.

\begin{figure*}
    \centering
    \includegraphics[width=0.95\textwidth]{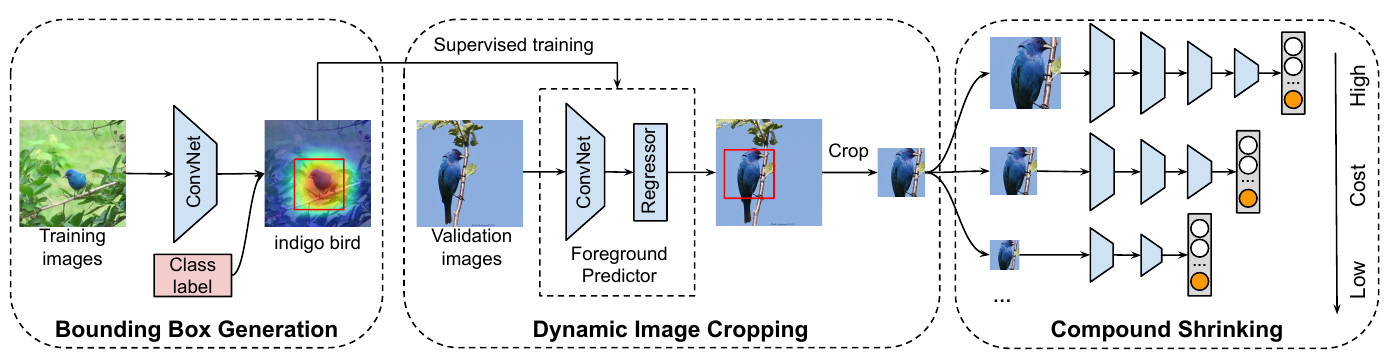}
    \caption{The overview of \textit{SmartScissor} \cite{kong2022smart}.}
    \label{fig:framework}
\end{figure*}

The overview of \textit{SmartScissor} is plotted in Fig. \ref{fig:framework}. The general idea behind \textit{SmartScissor} is to propose a method that can effectively and efficiently identify objects in input images such that it can facilitate the inference with low-resolution images, and then a more effective compound shrinking (CS) method can be determined under a complexity budget. In \textit{SmartScissor}, we propose a dynamic image cropping (DIC) component to find the object of interest in an input image. 
DIC first efficiently localizes the most discriminative foreground of the input image with a lightweight foreground predictor, then the detected foreground region will be preserved and the redundant background will be discarded. 
DIC is capable of generating fine-cropped images with less spatial redundancy, i.e., low resolution, thereby improving the inference accuracy even under low-resolution settings. 
The success of DIC attributes to the following two factors.  
\begin{figure}[!htbp]
    \centering
    \includegraphics[width=0.47\textwidth]{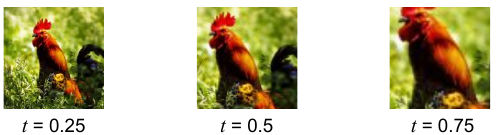}
    \caption{By applying different values of the salience threshold $t$, we can obtain different cropped images. The larger the threshold value, the more radical the cropping \cite{kong2022smart}.}
    \label{fig:threshold}
\end{figure}
\begin{itemize}
    \item \textbf{Data}: for classification datasets like ImageNet, there is no out-of-the-box position annotation for the foreground object. Moreover, the position of the foreground object varies in different images, which makes it difficult to efficiently localize the foreground object. To address this limitation, we have a bounding box generation before DIC, where we use Grad-CAM \cite{selvaraju2017grad} to automatically generate the position annotations. Grad-CAM is deployed to generate a salience map for each image, where a well-trained CNN (e.g. ResNet50) is applied. Then, the bounding box is gradually shrunk and determined based on a threshold $t$. Figure \ref{fig:threshold} shows the cropped images under different $t$, while Table \ref{tab:threshold} shows the different accuracy under different $t$.
    \item \textbf{Light-weight predictor}:  Although the foreground predictor we want is similar to object detectors and dozens of object detectors have been proposed, such as SSD and Faster R-CNN, these detector architectures are completely inapplicable to our task due to their high complexity. Applying the existing object detector will undermine the benefit of model compression, and an object detector may be more complex than a CNN model itself. We thus design a lightweight foreground predictor, whose detailed architecture is summarized in Table \ref{tab:arch}. It consists of several residual bottleneck blocks \cite{he2016deep} and a fully connected layer as the single-box regressor. 
A residual bottleneck contains two convolutional layers with 1$\times$1 kernels and one convolutional layer with 3$\times$3 kernels in the middle. 
The computational cost mainly results from the 3$\times$3 convolutional layer. Therefore, to reduce the cost and accelerate the predictor, we only stack two residual bottleneck blocks in each stage, and each block is only equipped with a small number of channels. 
The proposed predictor only contains 0.27M parameters and 0.09B MACs, which are negligible compared to popular object detectors (e.g., Faster R-CNN with 134.7M (499$\times$) parameters and 15.1B (167.8$\times$) MACs) in terms of model complexity. Moreover, the small overhead of the foreground predictor will be mitigated by CS.
\end{itemize}

\begin{table}[tbp]
  \centering
    \begin{tabular}{l|c|c|c|r}
    \hline
    \textbf{Model} & \textbf{$\#$Params} & \textbf{$\#$MACs} & \textbf{\textit{t}} & \textbf{Acc@1 } \\
    \hline
    \multirow{4}[2]{*}{ResNet50} & \multirow{4}[2]{*}{25.6 M} & \multirow{4}[2]{*}{4.1 B} & 0 (Baseline) & 76.02 $\%$ \\
        & &  & 0.25  & 76.45 $\%$ \\
        & &  & 0.5   & \textbf{76.88 $\%$} \\
        & &  & 0.75  & 76.32 $\%$ \\
    \hline
    \end{tabular}%
    \caption{The impact of using different salience thresholds on prediction accuracy. The model is trained and evaluated on ImageNet-1K. $t=0$ means using the original images without Grad-CAM cropping \cite{kong2022smart}.}
  \label{tab:threshold}%
\end{table}%

Once the predictor is trained, it can be directly applied to different classification backbones without any extra training overhead. During inference, the trained predictor will quickly localize the foreground object of an input image and generate a finely cropped image, which allows CNN models to predict the input image at lower resolution, thus significantly reducing computational cost.

\begin{figure}[tbp]
    \centering
    \includegraphics[width=0.47\textwidth]{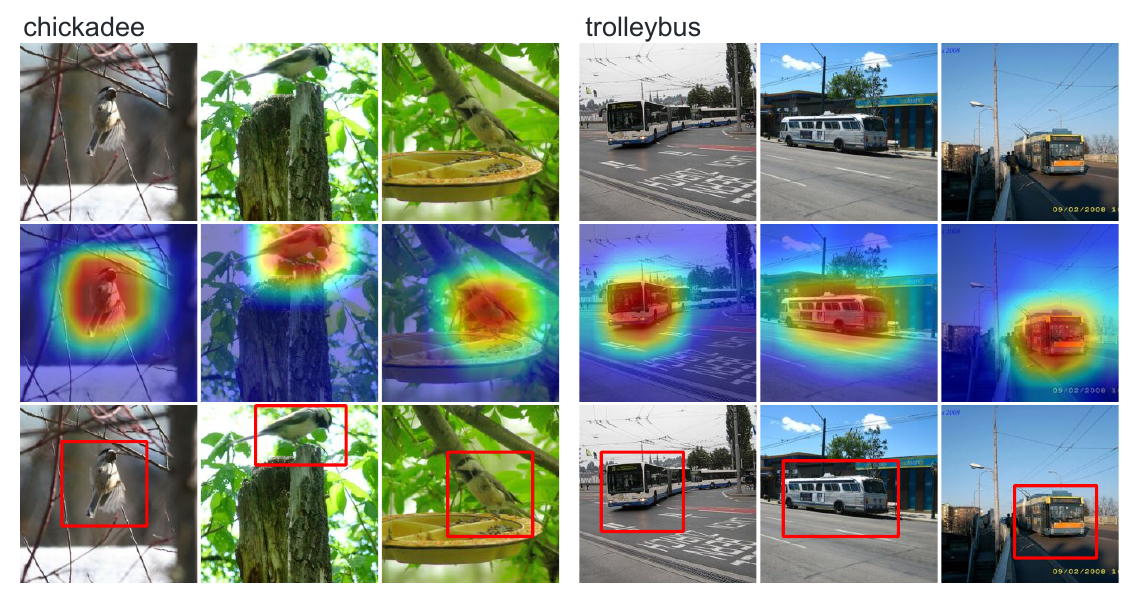}
    \caption{The bounding boxes generated with the salience threshold $t=0.5$, which accurately localize the key object in each image \cite{kong2022smart}.}
    \label{fig:grad-cam}
\end{figure}

\begin{table}[!thbp]
  \centering
    \begin{tabular}{lllll}
    \hline
    \multicolumn{1}{c||}{\textbf{Stage}} & \multicolumn{1}{c||}{\textbf{Block}} & \multicolumn{1}{c||}{\textbf{Resolution}} & \multicolumn{1}{c||}{\textbf{\#C}} & \multicolumn{1}{c}{\textbf{\#L}} \\
    \hline
    \multicolumn{1}{c||}{1} & \multicolumn{1}{c||}{Conv 3$\times$3} & \multicolumn{1}{c||}{224 $\times$ 224} & \multicolumn{1}{c||}{16} & \multicolumn{1}{c}{1} \\
    \multicolumn{1}{c||}{2} & \multicolumn{1}{c||}{Residual Bottleneck} & \multicolumn{1}{c||}{112 $\times$ 112} & \multicolumn{1}{c||}{16} & \multicolumn{1}{c}{2} \\
    \multicolumn{1}{c||}{3} & \multicolumn{1}{c||}{Residual Bottleneck} & \multicolumn{1}{c||}{56 $\times$ 56} & \multicolumn{1}{c||}{32} & \multicolumn{1}{c}{2} \\
    \multicolumn{1}{c||}{4} & \multicolumn{1}{c||}{Residual Bottleneck} & \multicolumn{1}{c||}{28 $\times$ 28} & \multicolumn{1}{c||}{32} & \multicolumn{1}{c}{2} \\
    \multicolumn{1}{c||}{5} & \multicolumn{1}{c||}{Residual Bottleneck} & \multicolumn{1}{c||}{14 $\times$ 14} & \multicolumn{1}{c||}{64} & \multicolumn{1}{c}{2} \\
    \multicolumn{1}{c||}{6} & \multicolumn{1}{c||}{Pooling \& Linear} & \multicolumn{1}{c||}{7 $\times$ 7} & \multicolumn{1}{c||}{4} & \multicolumn{1}{c}{1} \\
    \hline
    \multicolumn{5}{l}{\#Params: 0.27M} \\
    \multicolumn{5}{l}{\#MACs: 0.09B} \\
    \hline
    \end{tabular}%
    \caption{The architecture of the proposed box predictor. \#C denotes the number of channels and \#L denotes the number of layers \cite{kong2022smart}.}
  \label{tab:arch}%
\end{table}%

The proposed DIC underpins the low-resolution prediction. Now, we can look at the model complexity, width and depth. 
As shown in EfficientNet \cite{tan2019efficientnet}, CS that jointly compresses the three dimensions of a model promises higher accuracy over single-dimension compression. 
The key of CS is to calculate a shrinking coefficient for all dimensions according to their trade-off between accuracy and model complexity. 
Given the rule of thumb in CNN design, the more complex a model is, the more likely it is to achieve higher accuracy. 
With a number of MACs as the constraint or target in \textit{SmartSissor}, it is favorable to have a \textit{just-enough} shrinking, i.e., the number of MACs of the model compressed by CS can be approximately equal to the target number. 
Intuitively, shrinking different dimensions has different impacts on accuracy and model complexity. 
The shrinking coefficient needs to strike a good balance among the three dimensions.

EfficientNet exploits a time-consuming and computationally expensive grid search to determine the coefficient.
To calculate the shrinking coefficient efficiently, we first quantify the trade-off of each dimension between accuracy and model complexity. 
Here we use MACs as the metric to measure the cost of models, because all three dimensions are related to the MACs of a model while only the depth and width can affect the model parameters. 
Given a MACs budget $\mathcal{M}$, we first obtain the accuracy drops resulting from separately shrinking different dimensions, which can be represented as:
\begin{equation}
\label{eq:acc}
    \Delta A_s(\mathcal{M}) = A_0 - A_s(\mathcal{M})
\end{equation}
where $s\in\{d, w, r\}$ represents the shrunk dimension, $A_s(\mathcal{M})$ denotes the accuracy of the shrunk model, and $A_0$ denotes the accuracy of the original model. Based on our empirical analysis, we design the following equation to determine the shrinking coefficient for each dimension:
\begin{equation}
\label{eq:coeff}
    \mathcal{C}_s(\mathcal{M}) = \frac{\sqrt[3]{\Delta A_d(\mathcal{M}) \cdot \Delta A_w(\mathcal{M}) \cdot \Delta A_r(\mathcal{M})}}{\Delta A_s(\mathcal{M})}
\end{equation}
where $\mathcal{C}_s(\mathcal{M})$ denotes the shrinking coefficient of the dimension $s$ ($s\in\{d, w, r\}$). Through Eq.~(\ref{eq:acc}) and Eq.~(\ref{eq:coeff}), we are able to efficiently calculate the coefficients once we obtain the accuracy degradation of the three dimensions in the given MACs regime. 

\begin{figure}
    \centering
    \includegraphics[width=0.48\textwidth]{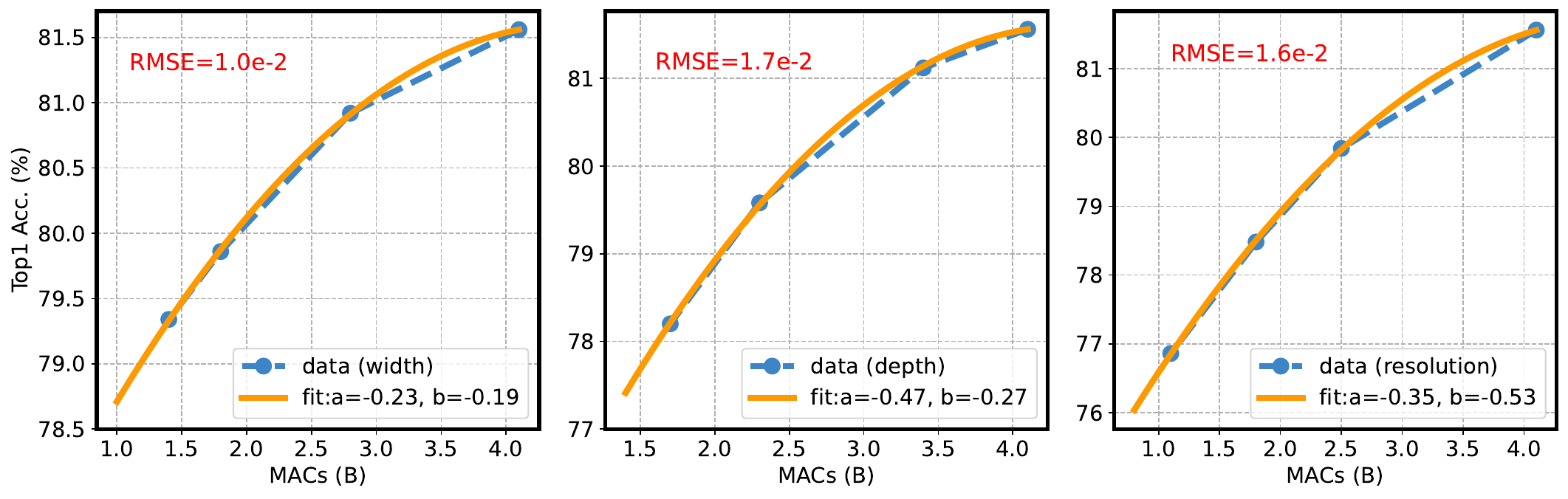}
    \caption{The actual accuracy (blue dotted line) and the estimated accuracy (yellow line) over MACs by separately shrinking the three dimensions. The low root mean square error (RMSE) indicates that the accuracy estimator can well fit existing data \cite{kong2022smart}.}
    \label{fig:dimensions}
\end{figure}

However, the training cost of calculating the accuracy drop is still non-negligible. To mitigate the training overhead, we propose a dimension-wise accuracy estimator to quickly estimate the accuracy of the compressed models and calculate the accuracy degradation resulting from shrinking different dimensions in the given MACs regime. 
First, we sample a couple of models with different MACs by separately shrinking the three dimensions. 
As demonstrated in Fig. \ref{fig:dimensions}, the accuracy distribution of the three dimensions along MACs can be well fitted by a quadratic polynomial. 
Therefore, we design a simple yet effective polynomial estimator to predict the accuracy with respect to the target MACs $\mathcal{M}$. The estimator can be formulated as follows:
\begin{equation}
    A_s(\mathcal{M}) = a_s (\mathcal{M} - \mathcal{M}_0)^2 + b_s (\mathcal{M} - \mathcal{M}_0) + A_0 
\end{equation}
where $\mathcal{M}_0$ is the MACs of the original model. $a_s$ and $b_s$ are the hyperparameters to fit for dimension $s$ ($s\in\{d, w, r\}$). 
Figure \ref{fig:dimensions} shows that the proposed estimator can well fit the existing data. 
Due to the simple and intuitive design of the estimator, we only need to sample and train very few models to train the estimator, and this cost is a one-time cost. With the accuracy estimator established, we are capable of directly calculating the accuracy drop and subsequently the shrinking coefficient across a wide range of MACs regimes. 
As a result, the cost of determining the coefficients is significantly reduced compared to directly training models to obtain the coefficients. We have conducted extensive experiments on different datasets in comparison with several state of the arts, where \textit{SmartSissor} can achieve higher accuracy with lower computational complexity \cite{kong2022smart}.


\subsection{LightNAS}
\label{sec:lightnas}

\begin{figure}[t]
    \begin{center}
    \includegraphics[width=1.0\columnwidth]{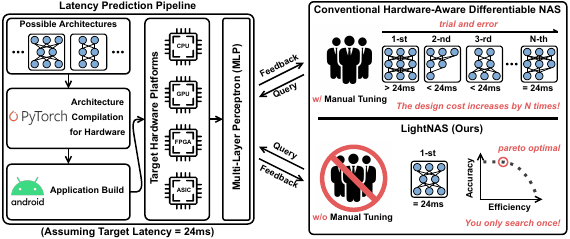}
    \end{center}
    \caption{An intuitive illustration of the proposed LightNAS framework \cite{luo2022lightnas}.}
    \label{fig:overview-lightnas}
\end{figure}

So far, we have discussed two frameworks that are able to optimize complex and redundant CNN models for edge systems. Another method to generate effective and efficient models for edge systems is to design a model from scratch, like MobileNet, ShuffleNet, etc, which best fit edge platforms. 
These models are hand-crafted by experienced ML practitioners based on their rich skills and knowledge. 
Recently, neural architecture search (NAS) has become an emerging technique to replace time-consuming hand-crafted design and automate the design of competitive CNNs. 
The well-established NAS methods can be divided into reinforcement learning methods, evolutionary methods, and gradient-based (a.k.a., differentiable) methods. 
Among them, differentiable NAS has started to take the stage of NAS research, thanks to its promising search efficiency. 
In this section, when referring to NAS, we mean differentiable NAS. 

To design hardware-efficient models for resource-limited edge systems, several hardware-aware NAS methods \cite{liu2022bringing} have been developed, which typically leverage the hardware performance metrics (e.g., latency and energy) to guide the search process for hardware-efficient models. The optimization objective can be formulated as follows:
\begin{equation}
    \small
    \mathop{\mathrm{minimize}}_{\alpha} \,\, \mathcal{L}_{valid}(w^*(\alpha), \alpha) + \lambda \cdot LAT(\alpha)
    \label{eq:fbnet-objective}
\end{equation}
where $\alpha$ represents a specific architecture generated by NAS, and $w^*$ denotes the weights of architecture $\alpha$. $LAT(\cdot)$ is the latency, and $\lambda \geq 0$ is a constant to control the trade-off magnitude between accuracy and latency. Generally, NAS is to find an architecture $\alpha$ that can minimize the loss function of $\mathcal{L}$ and latency regularization.

\begin{figure}[t]
    \begin{center}
    \includegraphics[width=1.0\columnwidth]{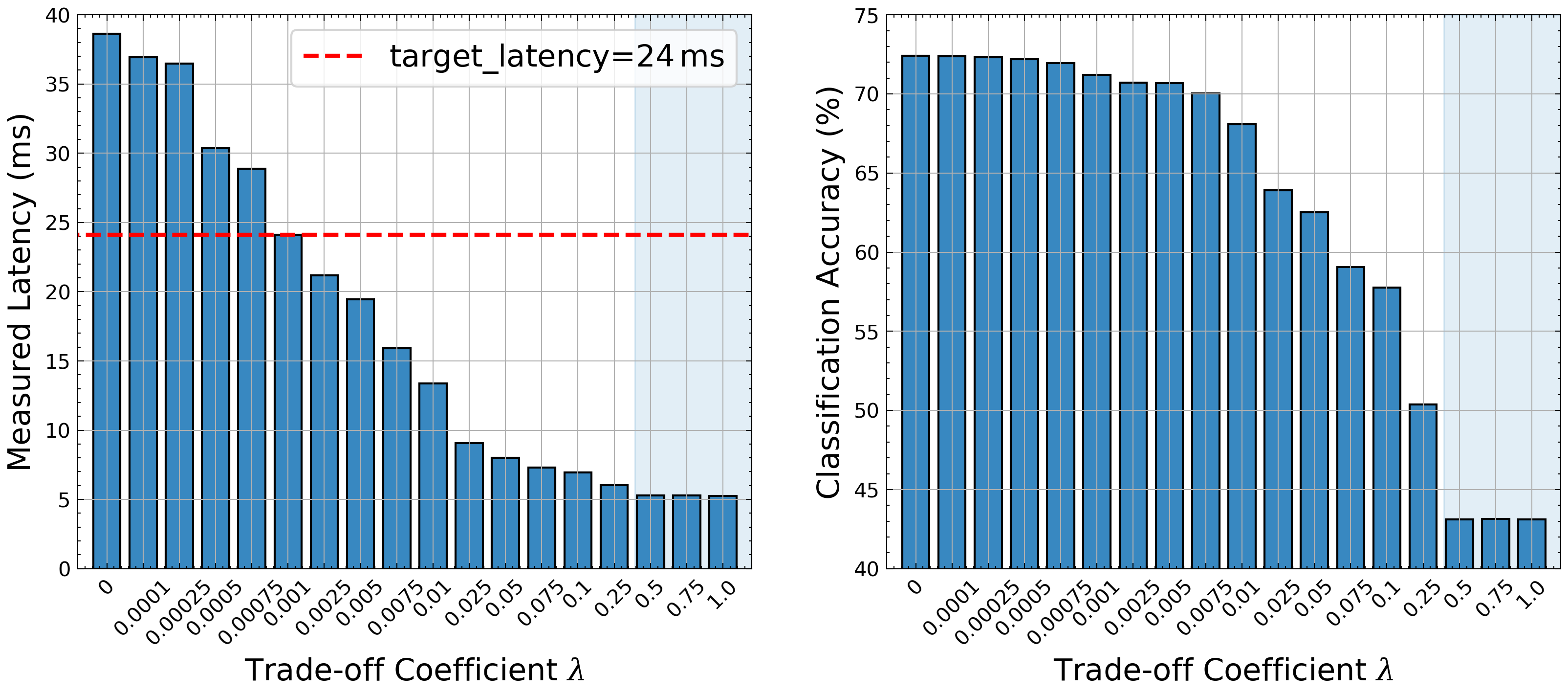}
    \end{center}
    \caption{Illustration of the architecture search results under $\lambda \in [0, 1]$ \cite{luo2022lightnas}.}
    \label{fig:tradeoff-lightnas}
\end{figure}

$\lambda$ in Eq.~(\ref{eq:fbnet-objective}) plays an unnoticed role in hardware-aware NAS. 
Similar to \textit{ZeroBN} and \textit{SmartScissor}, NAS expects to find the most complex model that can meet our performance requirement. 
To search an optimal model for an edge platform satisfying a given latency constraint, 
 $\lambda$ may need to be tuned for several or many rounds so that the searched model satisfies the specified hardware performance constraint. 
To show this, we leverage FBNet \cite{wu2019fbnet} to repeat a plethora of architecture search experiments under different settings of $\lambda \in [0, 1]$. As shown in Fig.~\ref{fig:tradeoff-lightnas}, $\lambda$ is able to trade off between accuracy and latency, which, unfortunately, is quite challenging to tune. As a result, to find the required architecture around the specified latency, the procedure has to repeat multiple search experiments (empirically 10), significantly increasing the search cost by $10\times$ times.

To overcome such limitations, we propose a novel hardware-aware NAS framework, dubbed \textit{LightNAS} \cite{luo2022lightnas}. 
\textit{LightNAS} aims to find the required architecture that satisfies a specified latency within one single search as shown in Fig.~\ref{fig:overview-lightnas} (i.e., \textit{you only search once}). The optimization objective of \textit{LightNAS} is similar to Eq.~(\ref{eq:fbnet-objective}) with a small modification as follows:
\begin{equation}
    \small
    \mathop{\mathrm{minimize}}_{\alpha} \,\, \mathcal{L}_{valid}(w^*(\alpha), \alpha) + \lambda \cdot \left( \frac{LAT(\alpha)}{T} - 1\right)
    \label{eq:objective-lightnas}
\end{equation}
where $T$ is the specified latency constraint. 
In \textit{LightNAS}, we use the hardware model discussed in Section \ref{model:hardware} to quickly evaluate the latency of architecture $\alpha$.
\textbf{Different from previous NAS methods, we set $\lambda$ in Eq.~(\ref{eq:objective-lightnas}) as a learnable hyper-parameter instead of a tuneable constant}. 
Therefore, the tedious manual hyper-parameter tuning can be replaced with an efficient learning procedure. 
And, \textit{LightNAS} automatically learns the optimal hyper-parameter configuration for $\lambda$ during the search process, which maximizes the accuracy while satisfying the specified latency constraint $LAT(\alpha)=T$. 
For simplicity, we use $\mathcal{L}(w, \alpha, \lambda)$ to denote the objective defined in Eq.~(\ref{eq:objective-lightnas}). Subsequently, $w$ and $\alpha$ are updated with gradient descent, whereas $\lambda$ is optimized using gradient ascent: 
\begin{equation}
    \small
    \begin{cases}
        w^* = w - \eta_w \cdot \frac{\partial\mathcal{L}(w, \alpha, \lambda)}{\partial w}, \,\,
        \alpha^* = \alpha - \eta_{\alpha} \cdot \frac{\partial \mathcal{L}(w, \alpha, \lambda)}{\partial \alpha} \\ 
        \lambda^* = \lambda + \eta_{\lambda} \cdot \frac{\partial \mathcal{L}(w, \alpha, \lambda)}{\partial \lambda} = \lambda + \eta_{\lambda} \cdot \left(\frac{LAT(\alpha)}{T} - 1\right)
    \end{cases}
    \label{eq:parameter-update}
\end{equation}
where $\eta_{w}$, $\eta_{\alpha}$, and $\eta_{\lambda}$ are the learning rates of $w$, $\alpha$, and $\lambda$, respectively. 
After demonstrating \textit{what} the proposed method is, we then analyze \textit{why} \textit{LightNAS} searches for an architecture $\alpha$ with $LAT(\alpha)=T$. 
$\lambda$ can adjust the complexity of the searched architecture, thereby affecting its latency.
A larger $\lambda$ derives the architecture with lower latency, whereas a smaller $\lambda$ generates the architecture with higher latency as shown in Fig.~\ref{fig:tradeoff-lightnas}. Then, during the search procedure, there are 2 possibilities.
\begin{itemize}
    \item $LAT(\alpha)>T$: The architecture does not meet the latency requirement, so the gradient ascent scheme of $\lambda$ increases $\lambda$ to reinforce the latency regularization magnitude;
    \item $LAT(\alpha)<T$: The architecture meets the latency requirement. However, if the architecture's latency is smaller than the required latency, we may not get \textit{the most complex model} which can maximize the attainable accuracy. Thus, the gradient ascent scheme then decreases $\lambda$ to diminish the latency regularization magnitude.
\end{itemize}
In both cases, the search engine will strive to make latency $LAT(\alpha)$ towards latency constraint $T$, i.e., $LAT(\alpha)=T$.
As a result, the search engine finally obtains an architecture $\alpha$ with $LAT(\alpha)=T$. Therefore, unlike previous hardware-aware NAS methods that require multiple trial-and-errors to find the desired architecture with latency $T$, \textit{LightNAS} only needs to search once, greatly improving the search efficiency.

\begin{figure}[t]
    \begin{center}
    \includegraphics[width=1.0\columnwidth]{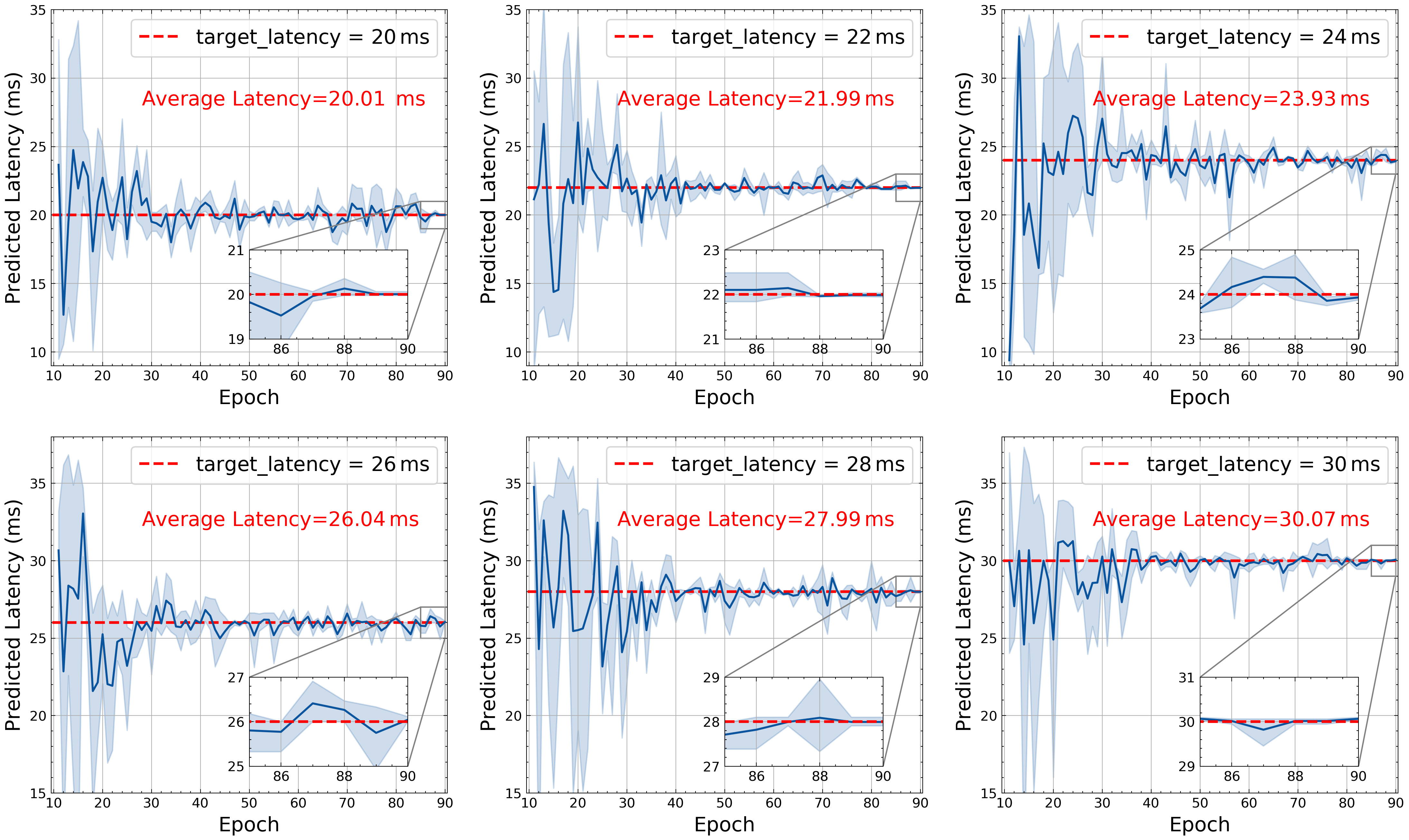}
    \end{center}
    \caption{Visualization of the search process under diverse latency constraints \cite{luo2022lightnas}.}
    \label{fig:latency-search-lightnas}
\end{figure}

\begin{figure}[t]
    \begin{center}
    \includegraphics[width=1.0\columnwidth]{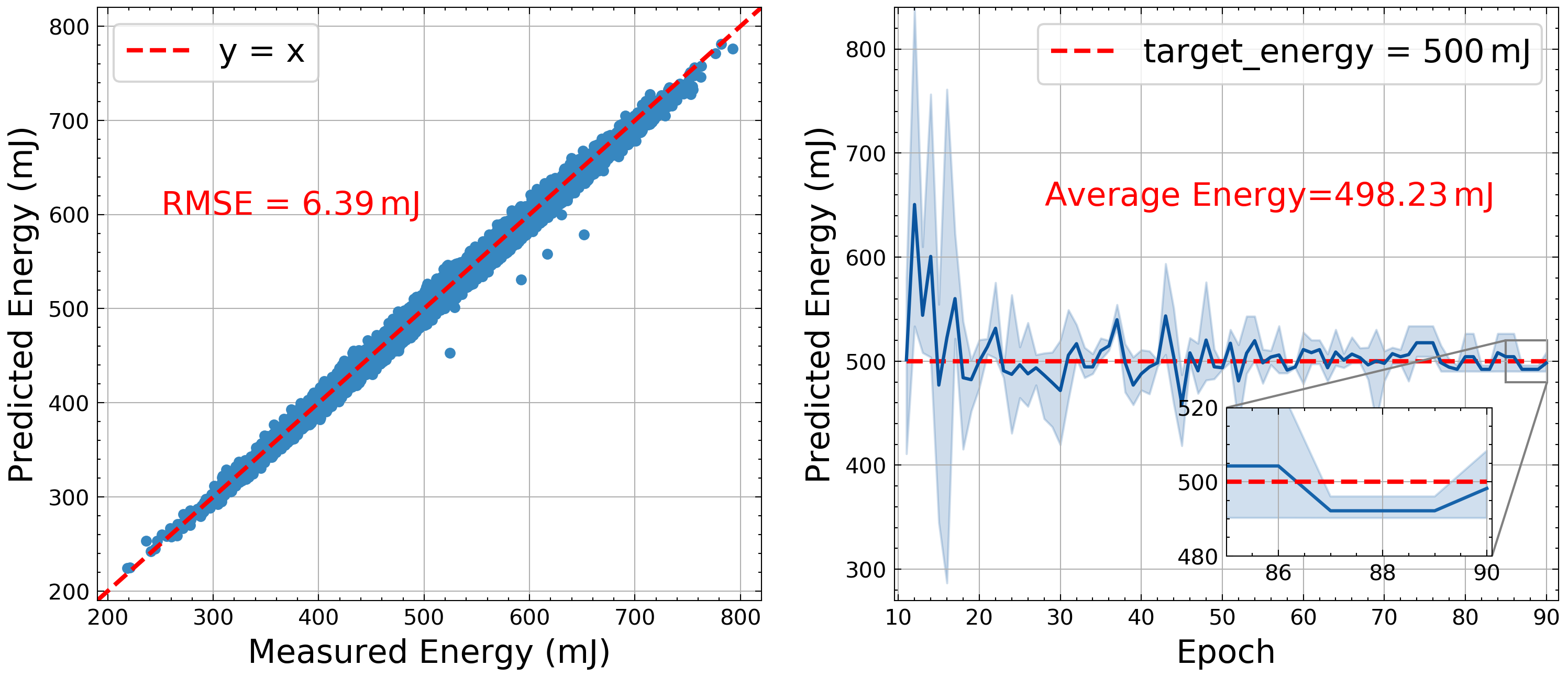}
    \end{center}
    \caption{Illustration of the generality of LightNAS to energy-critical search \cite{luo2022lightnas}.}
    \label{fig:energy-search-lightnas}
\end{figure}

\textbf{Results:} 
To demonstrate the effectiveness of \textit{LightNAS}, we visualize the search process under diverse latency constraints for a representative edge platform, Nvidia Jetson Xavier, in Fig.~\ref{fig:latency-search-lightnas}.  
The results clearly show that \textit{LightNAS} is able to search for the required architecture that strictly satisfies the specified latency constraint in one single search, where the latency of the searched architecture gradually converges to the required one. In addition, we also find that \textit{LightNAS} can be easily extended to deal with energy-critical tasks shown in Fig.~\ref{fig:energy-search-lightnas}, where we use the same method to model energy and replace the latency regularization with a new energy regularization. 
Experimental results clearly show the effectiveness of \textit{LightNAS} over previous state-of-the-art NAS methods \cite{luo2022lightnas}.

\section{Challenges}
\label{section:challenges}
In this section, we discuss some possible and unaddressed challenges in the field of edge intelligence especially in terms of hardware-aware design.

\subsection{Architecture-Aware Modelling}
The hardware modelling techniques discussed in Section \ref{section:back} and most literature \cite{liu2022bringing} feature a '\textit{black-box}' fashion, 
i.e., the modelling is unaware of what happens in the hardware and why different models or architectures perform so differently on the same hardware or on the same CNN model. 
Some work starts to look at how CNN computations are mapped into a underlying hardware and to analyze the performance of a model from hardware's perspective, like Timeloop \cite{parashar2019timeloop}. Many factors can affect the performance from hardware's perspective:
\begin{itemize}
    \item What are the key features of a target accelerator? Like, the number of processing units, the processing unit type, the underlying architecture, etc.
    \item What kind of underlying interconnect network does the accelerator deploy \cite{nabavinejad2020overview}? 
    \item How are computations mapped to the targeting accelerators? In another word, what is the dataflow of the target accelerator?
\end{itemize}
Although black-box modelling techniques can achieve relatively good performance in terms of prediction accuracy, they require to collect a huge amount of data. 
Also, the black box methods may impede the portability of the built model from one hardware platform to another. 
The procedure of '\textit{collecting-training-deployment}' has to be conducted for every new hardware. 
Moreover, if a new CNN architecture, new layers or new kernels are introduced, the existing prediction model may need to repeat the '\textit{collecting-training-deployment}'. 
This is mainly due to the lack of understanding how the target hardware works with CNN models. 
If some analytical methods based on hardware analysis can be integrated, a more explainable model is possible. 
For example, an analytical model for the processing unit may explain the reason why a newly introduced operator performs in a certain way and a communication analysis method can explain the rationale of data movements between layers or processing units \cite{nabavinejad2020overview}. 
Such architecture-aware models should be more precise, have lower training overhead, and provide better portability. 

\subsection{Unified Integration}
The methods discussed in Section \ref{section:methods} show different ways to optimize and design CNNs for the edge. 
We can either optimize an existing model or design a new model for a hardware. 
However, there is no evidence or conclusion which one is better or in which scenario a specific method should be deployed. 
This in turn leaves a new design decision for practitioners. If a task, a dataset, and an edge platform are given, a question may be immediately raised for engineers: which method should they use to implement the task on that edge platform? Besides performance and accuracy, many factors also have to be taken into account like training cost and portability. 
It thus would be desirable to have a methodological framework to integrate different design and optimization methods, then having a unified framework, in which different methods can be quantitatively compared in order to help practitioners to select the optimal method for their own tasks according to their requirements and consideration.

\subsection{Beyond CNN}
Most of works in edge intelligence focus on CNN-like models \cite{liu2022bringing}, because the majority of applications at the edge are vision-based. Prior to 2021, CNNs held a dominant position in the field. 
However, since the introduction of transformer in computer vision was first proposed in 2021, the landscape has undergone a significant shift. 
Moreover, generative AI models with strong capability are a new wave, like stable-diffusion models and ChatGPT models. These models are much more complex than CNN models, up to hundred billion parameters, and also collect more sensitive data from users, e.g., some personal information for generated images or dialogues, so they may be subject to more rigorous data protection. Thus, such models are expected to be increasingly implemented on edge systems and executed offline without the need to access remote servers. 
New models with architectures different from CNNs lead to new challenges for edge intelligence engineers. The concepts and insights we obtain for CNNs may still apply, but these methods need to undergo significant modifications or some new methods should be proposed for these emerging models.

\section{Conclusions}
\label{section:concl}
It is envisioned that more diverse AI applications will emerge and be integrated into our daily life in the near future. 
Edge as the new computing platform close to sensors and actuators has been gradually becoming one important hardware platform for AI applications. We briefly showcase three hardware-aware methods and conclude this paper with some existing challenges in edge intelligence research. We hope this paper can bring a new perspective to the community.
\section*{Acknowledgement}
This study is partially supported under the RIE2020 Industry Alignment Fund – Industry Collaboration Projects (IAF-ICP) Funding Initiative, as well as cash and in-kind contribution from the industry partner, HP Inc., through the HP-NTU Digital Manufacturing Corporate Lab (I1801E0028). This work is also partially supported by Nanyang Technological University, Singapore, under its NAP (M4082282/04INS000515C130).

\bibliographystyle{ieeetr}
\bibliography{reference}

\end{document}